\documentclass[aps,prb,showpacs,groupedaddress,twocolumn]{revtex4}
\usepackage{graphicx}
\usepackage{latexsym}
\usepackage{stmaryrd}

\begin{document}

\title{Numerical study of the topological Anderson insulator in HgTe/CdTe quantum wells}

\author{Hua Jiang$^{1}$, Lei Wang$^{1}$,  Qing-feng Sun$^{1}$, and X. C. Xie$^{2,1}$}

\address {\ $^1$Beijing National Lab for Condensed Matter Physics and
Institute of Physics, Chinese Academy of Sciences, Beijing 100190,
China;\\
$^2$Department of Physics, Oklahoma State University, Stillwater,
Oklahoma 74078 }

\date{\today}

\begin{abstract}
We study the disorder effect on the transport properties in the
HgTe/CdTe semiconductor quantum wells. We confirm that at a
moderate disorder strength, the initially un-quantized two
terminal conductance becomes quantized, and the system makes a
transition to the novel topological Anderson insulator (TAI).
Conductances calculated for the stripe and cylinder samples reveal
the topological feature of TAI and supports the idea that the
helical edge states may cause the anomalous quantized plateaus.
The influence of disorder is studied by calculating the
distributions of local currents. Base on the above-mentioned
picture, the phenomena induced by disorder in the quantum spin
Hall region and TAI region are directly explained. Our study of
the local current configurations shed further light on the
mechanism of the anomalous plateau.
\end{abstract}

\maketitle

\section{\protect\normalsize Introduction}
Quantum spin Hall effect (QSHE), which is proposed as a new class of
topological state of matter in two dimensions, has generated a great
deal of interest \cite{CLDay}. Contrary to the integer or the
fractional quantum Hall state, which is induced by a magnetic field
that breaks the time reversal symmetry, QSHE is caused by a strong
spin-orbit interaction that maintains the time reversal symmetry.
Kane and Mele proposed a $Z_{2}$ classification for this kind of the
new topological state\cite{CLKane}. For materials with spatial
inversion symmetry, the index could be easily computed as the
product of parities of the wave function at several high symmetry
points in the Brillouin zone. The $Z_{2}$ classification can be
generalized to three-dimensional systems as well \cite{Bi}.
Recently, topological insulators suitable for room temperature
applications are also predicted\cite{HaijunZhang}. The QSH state has
the helical edge states, namely, having two counter-propagating edge
states for the two opposite spin polarizations. The helical edge
states are stable against time-reversal conserving perturbations,
since backscattering processes need to connect the upper and lower
edges of the sample. The probability of backscattering is
exponentially suppressed as the sample width is increased. Recent
experiment \cite{MKonig} provides evidences of the QSHE in HgTe/CdTe
quantum well (QW) structures, as predicted
theoretically\cite{BABernevig}. The de-coherence effect in QSH
samples is also investigated.\cite{HJiang} Some interesting
questions emerge such as the response of a QSH state to the
disorder, and the interplay of the helical edge states and the bulk
states.

According to the celebrated scaling argument, in two dimensions the
bulk electron wave functions are localized in the presence of any
weak disorder \cite{AbrahansE}. But there are two known exceptions,
one is the two dimensional systems with a strong spin-orbital
coupling, and the other is the quantum Hall transition between
different plateaus\cite{FEvers}. With the discover of QSHE state in
two dimensions, study of the localization is in demand. Sheng et al
investigated the disorder effect in honeycomb lattice QSHE sample,
they found the QSHE phase is robust against weak
disorder\cite{LSheng}. Ono and Obuse \textit{et al} studied the
critical properties of the transition from a metal to a QSH state,
and they found the results to be somewhat
controversial\cite{Onoda,Obuse}.

Recently, Li\cite{JianLi} \textit{et al} studied the transport
properties of the HgTe/CdTe QWs in the presence of disorder, they
found as the increasing of the disorder strength, the initially
un-quantized conductance became quantized, \textit{i.e.} the sample
enters into the QSH state because of the disorder, so they named the
state as "topological Anderson insulator"(TAI). The newly anomalous
quantized conductance plateau is caused by the edge transport, which
is indirectly revealed by the unchanged plateau value along with
width variation for two-terminal calculation and quantized
transmission coefficient for four-terminal calculation. However, the
detailed mechanics of the edge transport is less clear.

In the present paper, we study the effect of disorder on the
electronic state of HgTe/CdTe QWs. We carry out Keldysh's
nonequilibrium-Green-function (NEGF) calculations based on a
four-band tight binding model. First, we perform the conductance
calculations for two different geometries. In the case of a stripe
geometry(see fig .1(a)), the presence of the helical edge states are
evident from the band spectrum, the TAI phase described in reference
12 appears. While for the cylindrical geometry, \textit{i.e.}
periodical boundary condition along $y$ direction(see fig. 1(b)),
there is no edge state, the bulk state is localized by disorder and
there is no quantized conductance. These results strongly support
the thesis that anomalous conductance plateau is due to the edge
transport and give a better understanding of the topological feature
of the TAI phase. Second, we obtain the distributions of the local
currents for the two-terminal strip samples with different chemical
potentials and disorder strengths. The evolvement of the local
current vector configurations gives rise to a direct demonstration
of the impurity influence in the HgTe/CdTe QWs. By analyzing these
local current configurations, the transport phenomena in both normal
QSH region and TAI region are clearly explained. Moreover, the
detailed results, such as the coexistence of the bulk and edge
states at the dip point (see Fig.6) and the bulk states assisted
backscattering obtained from the local current vector configurations
shed further light on the mechanism of the disorder induced edge
states for the TAI.

%

The rest of this paper is organized as follows. In Section II, we
introduce the effective tight-binding model. The formulas and
calculation method are also described. In Section III, the numerical
results and their discussions are presented. Finally, a conclusion
is given in Section IV.

\section{\protect\normalsize The Model and Method}

As a starting point, we introduce the effective Hamiltonian for
the HgTe/CdTe QWs with Anderson impurity in the tight-binding
representation. We consider a square lattice with four special
orbit states $|s,\uparrow \rangle $,$ |p_{x}+ip_{y},\uparrow
\rangle $ , $|s,\downarrow \rangle $, $
|-(p_{x}-ip_{y}),\downarrow >$ on each site. Here $\uparrow
$,$\downarrow $ denotes the electron spin. Through symmetry
consideration,
the effective Hamiltonian can be written
as\cite{BABernevig,CXLiu,MKonig,JianLi}

\begin{eqnarray}
H &=&\sum_{\mathbf{i}}\varphi _{\mathbf{i}}^{\dagger }\left(
\begin{array}{cccc}
E_{\mathbf{i}sa} & 0 & 0 & 0 \\
0 & E_{\mathbf{i}pc} & 0 & 0 \\
0 & 0 & E_{\mathbf{i}sb} & 0 \\
0 & 0 & 0 & E_{\mathbf{i}pd} \\
\end{array}
\right) \varphi _{\mathbf{i}}  \nonumber \\
&+&\sum_{\mathbf{i}}\varphi _{\mathbf{i}}^{\dagger }\left(
\begin{array}{cccc}
V_{ss} & V_{sp} & 0 & 0 \\
-V_{sp}^{\ast } & V_{pp} & 0 & 0 \\
0 & 0 & V_{ss} & V_{sp}^{\ast } \\
0 & 0 & -V_{sp}^{{}} & V_{pp} \\
\end{array}
\right) \varphi _{\mathbf{i}+\delta x}+h.c.  \nonumber \\
&+&\sum_{\mathbf{i}}\varphi _{\mathbf{i}}^{\dagger }\left(
\begin{array}{cccc}
V_{ss} & iV_{sp} & 0 & 0 \\
iV_{sp}^{\ast } & V_{pp} & 0 & 0 \\
0 & 0 & V_{ss} & -iV_{sp}^{\ast } \\
0 & 0 & -iV_{sp} & V_{pp} \\
\end{array}%
\right) \varphi _{\mathbf{i}+\delta y}+h.c.  \nonumber \\
\end{eqnarray}%
Here $\mathbf{i}=(ix,iy)$ is the site index, and $\delta x$ and $
\delta y$ are unit vectors along the $x$ and $y$ directions.
$\varphi _{
\mathbf{i}}=(a_{\mathbf{i}},c_{\mathbf{i}},b_{\mathbf{i}},d_{\mathbf{i}
})^{T} $ represents the four annihilation operators of electron on
the site $ \mathbf{i}$ with the state indexes $|s,\uparrow \rangle
$,$ |p_{x}+ip_{y},\uparrow \rangle $ , $|s,\downarrow \rangle $, $
|-(p_{x}-ip_{y}),\downarrow >$, respectively. The on-site matrix
elements satisfy
$E_{\mathbf{i}sa}=E_{s}+W_{\mathbf{i}sa},E_{\mathbf{i}pc}=E_{p}+W_{
\mathbf{i}pc},E_{\mathbf{i}sb}=E_{s}+W_{\mathbf{i}sb},E_{\mathbf{i}
pd}=E_{p}+W_{\mathbf{i}pd}$.
$W_{\mathbf{i}sa},W_{\mathbf{i}pb},W_{\mathbf{i }sc}$, and
$W_{\mathbf{i}pd}$ are on-site disorder energies uniformly
distributed in the range $[-\frac{W}{2},\frac{W}{2}]$ with the
disorder strength $W.$ $E_{s},E_{p},V_{ss},V_{pp},$ and $V_{sp}$ are
the five independent parameters that characterize the clean
HgTe/CdTe samples. It is clear that near the $\Gamma $ point the
lattice Hamiltonian (Eq. (1)) in k-representation can be reduced to
the continuous Hamiltonian in Ref.6 when we take $V_{sp}=-iA/2a$, $
V_{ss}=(B+D)/a^{2}$, $V_{pp}=(D-B)/a^{2}$, $E_{s}=C+M-4(B+D)/a^{2}$,
and $ E_{p}=C-M-4(D-B)/a^{2}$. Here $a$ is the lattice constant and
all the parameters $A,B,C,D,M$ can be controlled
experimentally\cite{MKonig}. Moreover, the Eq.(1) can be directly
obtained by discretizing spatial coordinates of the continuous
Hamiltonian using the substitution $ k_{x}\rightarrow
-i\frac{\partial }{\partial x},k_{y}\rightarrow -i\frac{
\partial }{\partial y}$.\cite{SDatta}

In this paper, we apply the model to two geometric devices. The
device a (see fig. 1(a)) is of stripe geometry, while the device b
(see fig. 1(b)) is of cylindrical geometry which can be obtained by
rolling the device $a$  into a tube. Without disorder(W=0), the
energy spectrum of such two geometric devices can be calculated by
diagonalizing Eq.(1) using the periodic boundary condition in x
direction \cite{XLQi,YHatsugai}. Next we investigate how the
disorder affects the transport properties of such systems. For both
devices a and b, the size of the central region is $L \times W$. For
convenience, we assume that the Anderson impurities only exist in
the (red) filled region and the temperature is zero.

In our simulations, a small external bias $V=V_{L}-V_{R}$ is
applied between the two terminals. With the help of the NEGF
method, the local current between neighboring sites $\mathbf{i}$
and $\mathbf{j}$ is calculated from the
formula\cite{APJauho,NNakamishi}
\begin{equation}
J_{\mathbf{i}\rightarrow \mathbf{j}}=\frac{2e^{2}}{h}Im[\sum_{\alpha
,\beta }H_{\mathbf{i}\alpha ,\mathbf{j}\beta }G_{\mathbf{j}\beta
,\mathbf{i} \alpha }^{n}(E_{F})](V_{L}-V_{R})
\end{equation}
where $V_{L},V_{R}$ are the voltages at the $Lead$-L,R. $\alpha
,\beta $ denote the state indices. $G^{n}(E_{F})=G^{r}\Gamma
_{L}G^{a}$ is electron correlation function with line width function
$\Gamma _{\alpha }(E_{F})=i[\Sigma _{\alpha }^{r}(E_{F})-(\Sigma
_{\alpha }^{r}(E_{F}))^{+}]$ , the Green function
$G^{r}(E_{F})=[E_{F}-H_{cen}-\Sigma _{L}^{r}(E_{F})-\Sigma
_{R}^{r}(E_{F})]^{-1}$ and the Hamiltonian in the central region
denoted as $H_{cen}$. The retarded self-energy $\Sigma
_{L}^{r},\Sigma _{R}^{r}$ due to the coupling to the lead-$L,R$ can
be calculated numerically \cite{DHLee}. Note that for Eq. (1), the
spin up($a_{\bf i},c_{\bf i}$) subsystem and spin down ($b_{\bf
i},d_{\bf i}$) subsystem are decoupled. The local current between
neighboring sites $\mathbf{i}$ and  $\mathbf{j}$ with spin index
$\sigma$ $J_{{\bf i}\rightarrow {\bf j}}^{\sigma}$  can also be
calculated from Eq.(2) by summing over only the state index with the
corresponding unitary subsystem. The current $J_{L}$ flowing through
the device is calculated by summing all the local current $J_{
\mathbf{i}\rightarrow \mathbf{i}+\delta x}$ for an arbitrary column.
After obtaining the current $J_{L}$ , the linear conductance
$G_{LR}$ is given by $G_{LR}=J_{L}/(V_{L}-V_{R})$. In addition, the
linear conductance can be directly obtained by $G_{LR}=Tr[\Gamma
_{L}G^{r}\Gamma _{R}G^{a}]$. The agreement between the two methods
gives strong confirmation of our analytical derivations and
numerical calculations.

\begin{figure}
\includegraphics[width=9cm,totalheight=6cm]{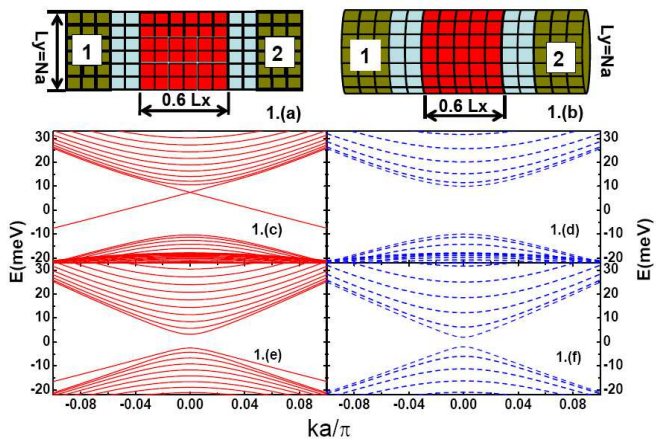}
\caption{(color online) (a) and (b)are the schematic diagrams for
two devices. The Anderson impurities only exist in the (red) filled
regions. (c) and (e) show the one-dimensional energy bands for
device a. The parameters are  $A=3.645eV\cdot {\AA}, B=-68.6eV\cdot
{\AA}^2, C=0.0meV, D=-51.2eV\cdot {\AA}^2$, and gap parameter
$M=-10meV$ (c), $M=2meV$ (e). (d) and (f) demonstrate the
one-dimensional energy bands for device b with the same parameters
as (c) and (e), respectively.}
\end{figure}

In the following numerical calculations, we choose the realistic
material parameters $A,B,C,D,M$ that arrived from the HgTe/CdTe
QWs.\cite{MKonig} The sample width (or circumference) is fixed to
$L_{y}=80a$ with the lattice constant $a=5nm$ . Since the model is
only valid in small $k$, we confine the Fermi energy within a
small region near the $\Gamma$ point. In the presence of disorder,
the conductance $G$, conductance fluctuation $\delta G$, the local
current $J_{{\bf i}\rightarrow {\bf j}}^{\sigma}$ etc., are all
averaged over up to 500 random configurations.

\begin{figure}
\includegraphics[width=8.1cm,viewport=43 125 750 550 clip]{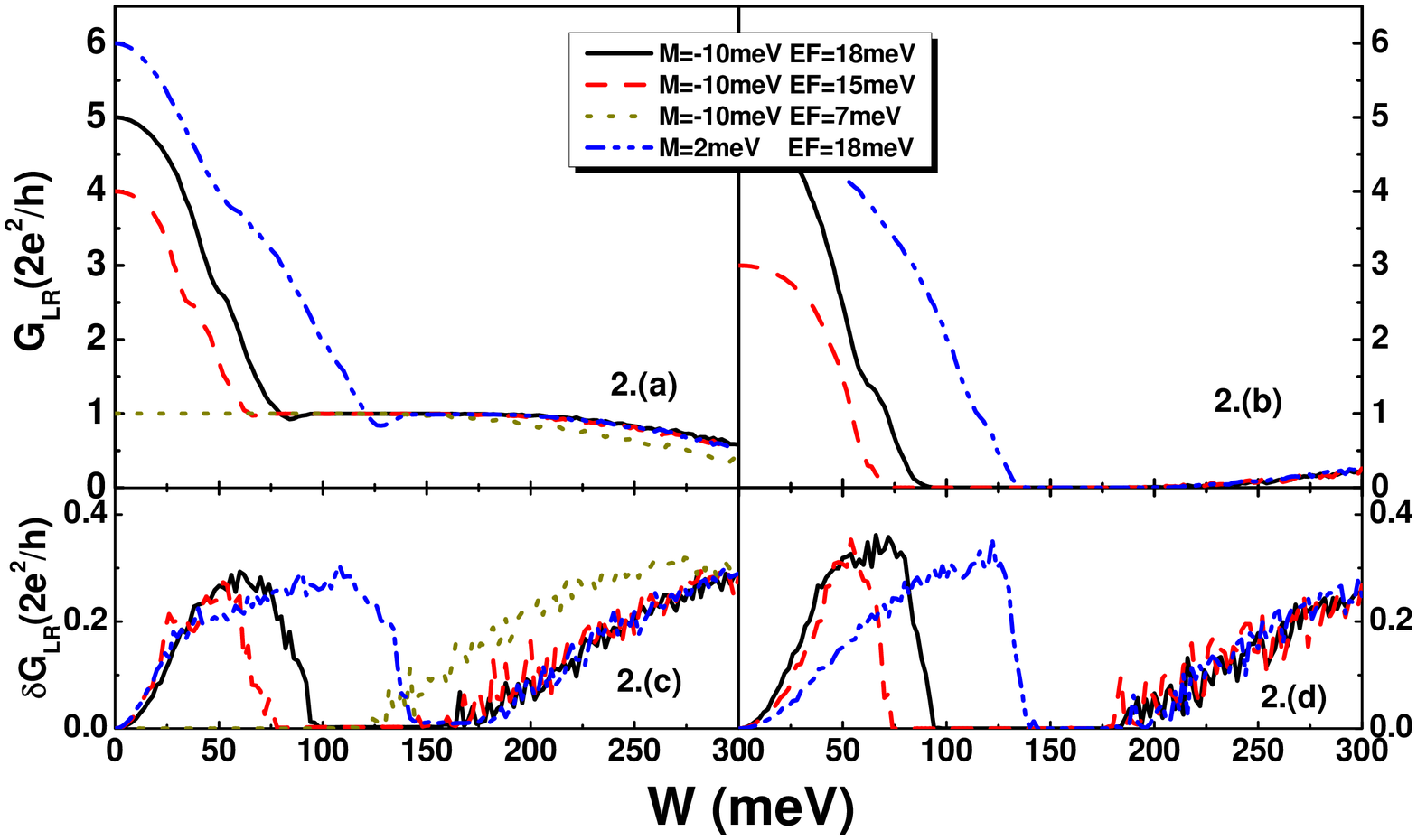}
\caption{ (color online) The conductance $G$ (a)(b) and
conductance fluctuation $\delta G$ (c)(d) vs disorder strength W
for different Fermi energy $E_{F}$ and gap parameter M. (a)(c) for
device a and (b)(d) for device b. The central region length
$L_{x}=200a$. Other parameters (A, B, C, and B) are the same as fig. 1(c)-1(f).
}
\end{figure}

\section{\protect\normalsize NUMERICAL RESULTS}

Let us first compare the two devices in geometry and topology. In
device $a$ with open boundary (fig.1(a)), there are two edges in the
y direction. Thus, the edge states can exist in such device. Since
the fig. 1(b) is a cylinder without edge, the edge states are
prohibited in the device b.  In figures 1(c) and 1(e), the band
structures of clean HgTe/CdTe QWs of are plotted. There exist a bulk
energy gap approximately of $2|M|$ in both figures. Moreover, there
are two degenerate bands (edge states) cross inside the gap for gap
parameter $M<0$ (fig. 1(c)). In contrast, the crossing bands
vanished when $M$ was tuned up to above zero (fig. 1(d)). These
results are in agreement with those of previous
studies\cite{BABernevig,XLQi}. In figures 1(d) and  1(f), we plot
the corresponding band structures in the cylindrical geometry. For
both gap parameter $|M|>0$ and $|M|<0$, though the degeneracy is
higher due to the enhanced symmetry, the bulk energy gap is nearly
unchanged and there is only tiny shift in the energy bands. However,
there is a big difference between the two samples, the edge states
which cross inside the gap vanish in the later one. The phenomenon
originates from the topology of the device, when the sample is
sufficient large, the bulk states are hardly affected by the
topology change at the edge while the edge states are totally
destroyed. In conclusion, rolling the sample from a strip to a
cylinder destroys the edge channels but maintain the bulk state
properties.

Next, we investigate how the transport properties are affected by
disorder. Fig .2(a) and 2(c) plot the conductance $G$ and $\delta G$
versus disorder strength for device a. When the system is in the
inverted regime ($M<0$) with the Fermi energy inside the bulk gap
($E_F=7meV$), for a range of disorder strength $W$, the two terminal
conductance is quantized without much fluctuations. Such observation
agrees with the previous result that the QSHE is robust against weak
disorder \cite{CLKane,LSheng}. However, when the chemical potential
is tuned up into the bulk band region near the $\Gamma$ point, no
matter $M>0$ or $M<0$, the new intriguing phenomena emerge. The
conductance $G$ decreases while the fluctuation $\delta G$ increases
when the disorder is first applied. When the disorder strength
continues to increase, instead of localization, the conductance
begins to increase to a quantized value ($2e^2/h$) and maintains at
this value for a certain range before eventually decreases.
Meanwhile, the conductance fluctuation $\delta G$ decreases to zero
and vanishes for the corresponding W. The anomalous conductance
plateau indicates that the sample becomes a topological insulator.
More importantly, different from the traditional topological
insulator (QHE etc), the quantized value is induced by the
impurities. This disorder induced anomalous conductance plateau was
discovered in a very recent work.\cite{JianLi}
Naturally, there exists a question: what is the mechanism that
causes such an anomalous plateau?

\begin{figure}
\includegraphics[width=9.0cm,viewport=29 130 580 780 clip]{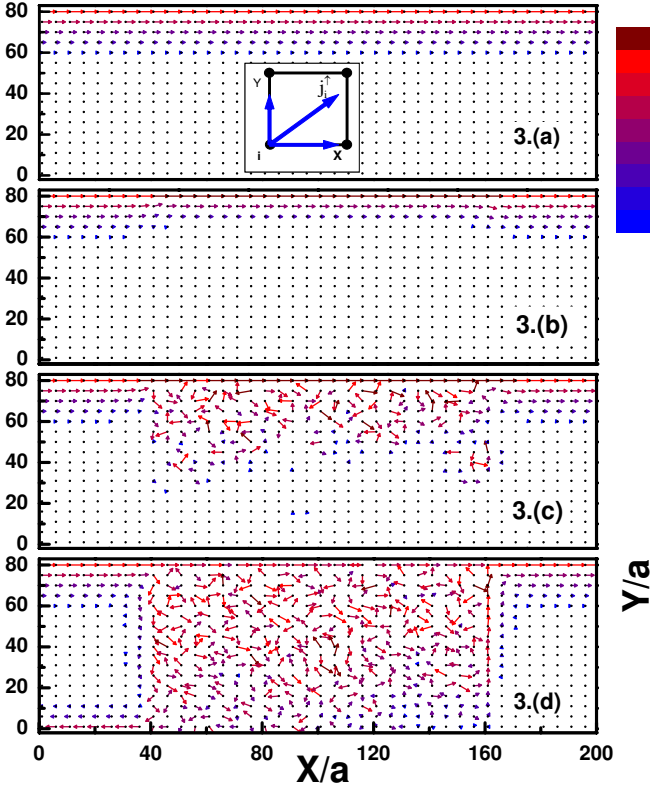}
\caption{ (color online) Configurations of the local current flow
vector for device a with Fermi energy $E_{F}=7meV$,M=-10meV, central
region size $L_{x}=200a,L_{y}=80a$ under disorder strength W=0 (a)
W=50meV(b) W=110meV (c) W=220meV (d). The inset in fig. 3(a) is the
schematic of local current flow vector. The vector direction
represent the local current flow direction and the vector length is
proportional to the logarithm of local current value.}
\end{figure}

To answer this question, we first examine $G$ and $\delta G$ as a
function of disorder strength for device b using the same parameters
as device a [see Figs. 2(b) and 2(d)]. Unlike the device a, the
transport properties follow the traditional Anderson metal-insulator
transition feature,\cite{FEvers} and the anomalous conductance
plateau is absent. For instance, the conductance monotonously
decreases to zero with increase of the disorder strength and the
critic disorder strength $W_{c}$ increases with raising of the Fermi
energy. Significantly, the metal-insulator transition point for
device b is roughly at the starting point of the anomalous plateau
for device a. Take $M=-10meV,E_{F}=18meV$ for example [see black
solid line in fig. 2(a) and 2(b)], the anomalous plateau sets up at
the disorder strength $W_{c}\simeq 93meV$ for device a and for that
threshold value the device b becomes an insulator($G\simeq 0,\delta
G\simeq 0$). As shown in fig .1, the bulk states in the device a and
the device b are the same. Only the edge states are completely
destroyed by rolling the device a into the device b. This gives a
direct evidence that the anomalous conductance plateau originates
from the edge states.

%

\begin{figure}
\includegraphics[width=9.4cm,viewport=29 167 600 782 clip]{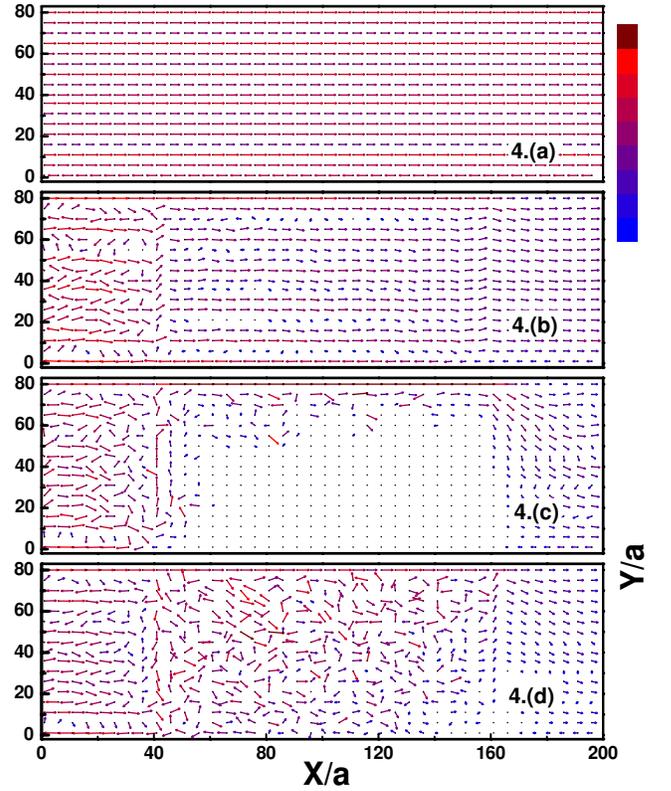}
\caption{ (color online) Configurations of the local current flow
vector for device a with the same sample sizes as for fig.3,
positive gap parameter $M=2meV$, Fermi energy $E_{F}=18meV$, and
disorder strength W=0 (a) W=100meV (b) W=150meV (c) W=250meV (d).}
\end{figure}

\begin{figure}
\includegraphics[width=10.2cm,viewport=16 153 655 794 clip]{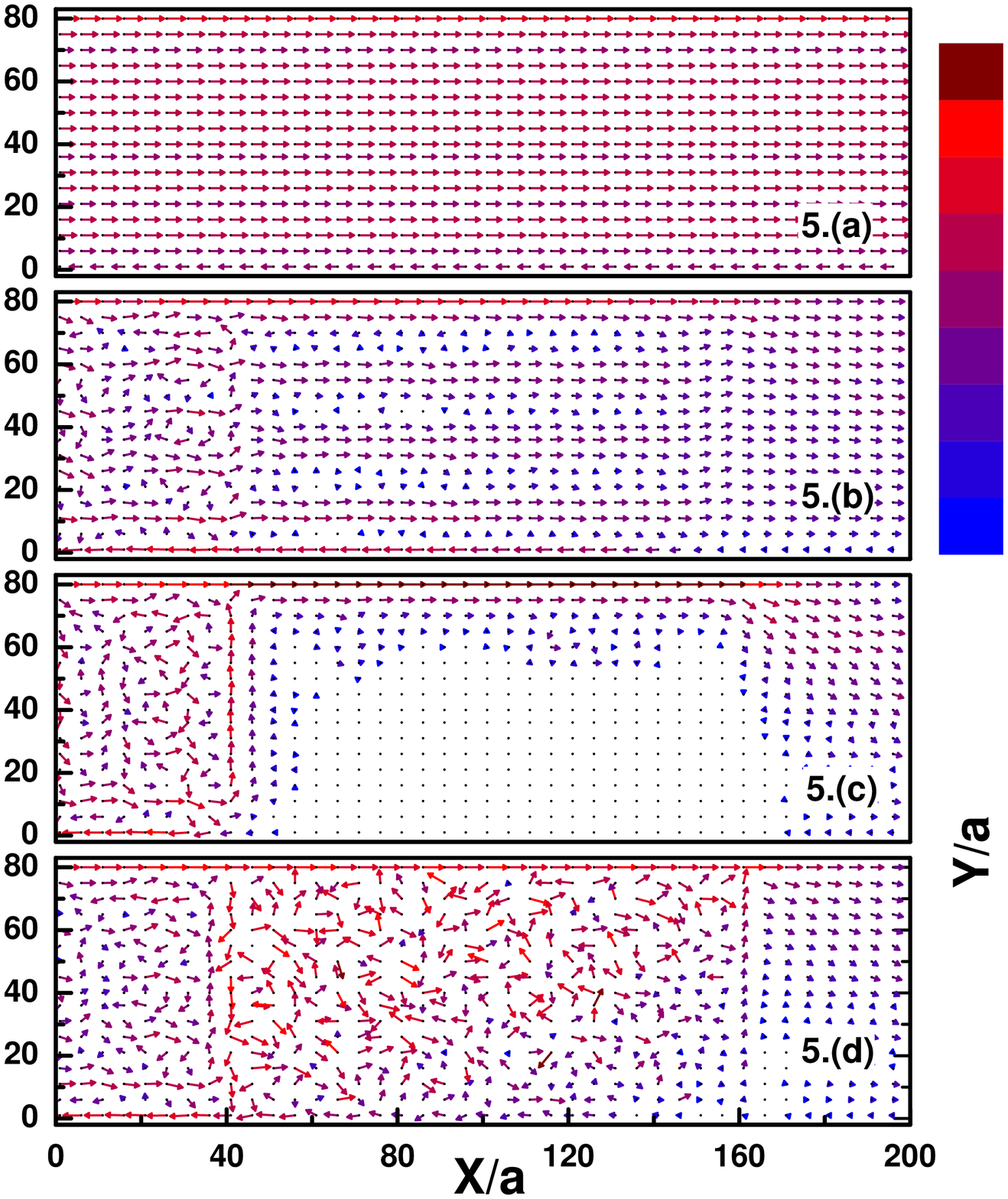}
\caption{ (color online) Configurations of the local current flow
vector for device a  with the same sample sizes as for fig .3,
negative gap parameter $M=-10meV$, Fermi energy $E_{F}=18meV$, and
disorder strength W=0 (a) W=65meV (b) W=100meV (c) W=250meV (d).}
\end{figure}

To get a better insight into the microscopic origin of the
conductance variations, we examine the disorder effect through the
local-current-vector-flow configurations. Due to the time reversal
symmetry, we only consider the spin up subsystem, the influence of
spin down subsystem can be directly obtained by time reversal
symmetry. Here, the local current flow vector on site {\bf i} is
defined as $J_{\bf i}^{\uparrow}=J_{\bf{i} \rightarrow
\bf{i+\delta x}}^\uparrow+J_{ \bf{i}\rightarrow \bf{i+\delta y
}}^\uparrow$.


In fig.3, the typical distributions of local currents for device a
in traditional QSHE region are plotted. For a clean sample (see
fig.3(a)), the local currents locate mainly on the upper edge and
their values decay exponentially towards the bulk. Surprisingly, the
small disorder initially makes the edge channel narrower (see fig.
3(b)). Though the mechanism is unclear, we note such phenomenon was
also observed recently by Chu \textit{et al}.\cite{Rui-LinChu} In
their paper, this narrowing effect was indirectly observed by the
decreasing of the oscillation period of the A-B ring, while in this
paper such effect is directly shown by the spatial distributions of
the local currents. When the Anderson disorder strength is getting
larger,
the local currents spread to the bulk and broaden the edge channels
again(see as fig. 3(c)). However, only when the disorder strength
exceeds the critical value $W_{c}$, the spread local-current flow
can reach the lower edge channels with different chirality, the
effective backscattering (as shown in the local current flow vector
located near the lower edge in the region $0<X<40a$ in fig .3(d))
can take place, leading to the reduction of the conductance between
the two terminals. These pictures explain why the traditional
quantized plateau is robust under weak disorder and how it is
destroyed in the strong disorder limit.


\begin{figure}
\includegraphics[width=8.7cm,viewport=35 89 580 560 clip]{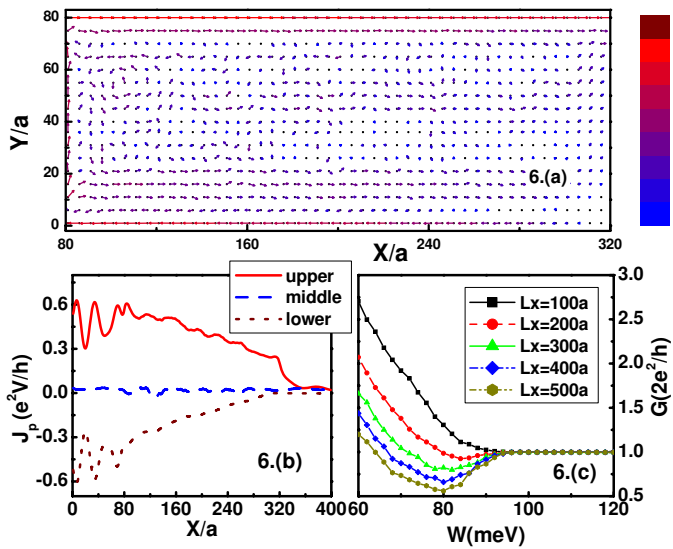}
\caption{ (color online) (a) Distribution of local currents in the
disorder region for device a with length $L_{x}=400a$ and
$L_{y}=80a$ at the dip point. (b) The position-related current
$J_{p}$ vs longitude axis x. $J_{p}$ is defined as the summation of
$j_{{\bf i}\rightarrow {\bf i}+\delta x}^{\uparrow}$ for four layers
in the corresponding region. The other parameters are the same as
(a). (c) The two-terminal conductance G vs. disorder strength W with
different sample lengths $L_{x}$. }
\end{figure}

Next, the Fermi energy is tuned to $E_{F}=18meV$, sitting slightly
above the bulk gap. A positive gap parameter $M=2meV$, for which
there is no helical edge states inside the bulk gap for the clean
sample, is chosen in the following simulations. The conductance $G$
vs. disorder strength $W$ shown in fig. 2(a) can be classified to
four regions (i) without disorder, (ii) before the anomalous
plateau, (iii) on the anomalous plateau, and (iv) after the
anomalous plateau. The typical configurations of local-current-flow
vector in such four regions are plotted in fig.4. For the clean
sample (see fig. 4(a)), the local current not only flow forward
along the upper edge, but
also uniformly flow forward in the bulk. 
While $W$ is increased from zero into the region (ii), the local
currents in the bulk of the disorder region become smaller and more
irregular in directions (see fig. 4(b)), which directly shows the
decline of the bulk transport and the localization of the bulk
states. Fig .4(c) demonstrates the distribution of the local
currents when the device is in the region (iii). The most
interesting phenomena exist in the disorder region $X\in
[40a,160a]$. The local currents in the bulk decline to zero while
the residual currents flow in the upper edge and flow without much
scattering. In addition, throughout the region (iii), the bulk
transport vanishes due to the localization of all the bulk states.
In contrast, the edge transport shows the same behavior as in the
traditional QSHE region: the local currents deeply spread into the
bulk without any effective backscattering with $W$ increasing until
it connected the opposite edge channels(see as fig. 4(d)).
Obviously, such edge transport will lead to the quantized plateau.
In the negative gap parameter situation, for example
$M=-10meV,E_{F}=18meV$,(see fig. 5) the evolvement of local current
configurations resembles that of $M>0$ situation. So far, the fig.
4(c) and fig. 5(c) are the most strong evidence directly showing
that the TAI is caused by the edge transport. In addition, these
plots provide a vivid microscopic picture demonstrating the
influence of disorder on TAI.

   Up to now, we have explained the majority phenomena emerged
in the TAI. Nevertheless, the transport properties in the region
(ii) of $G-W$ figure demands a detailed study for the following two
reasons. (i) The conductance is not monotonously decreasing with
increasing disorder strength but shows a dip feature prior to the
anomalous plateau. Obviously, at the dip point, the system is
neither in TAI phase nor in normal Anderson insulator phase due to
its nonzero and non-quantized conductance value. (ii) For both types
of gap parameter $M>0$ and $M<0$, the dip does exist prior to the
anomalous plateau and the conductance $G$ behaves similarly after
the dip. Thus, revealing the cause for the dip feature may help us
to understand the mechanism of the formation of the edge states.

In fig.6(a), the distribution of the local currents with the
disorder strength $W$ being fixed at the dip is plotted. The local
currents flow are larger close to the edges than in the bulk. The
predominant edge transport is clearly seen. In fig. 6(b) we plot the
position-related current $J_{p}$ versus longitude axis $x$. Where
$J_{p}$ is defined as the summation of $j_{{\bf i}\rightarrow {\bf
i}+\delta x}^{\uparrow}$ for four layers in the corresponding
region. For example, given sample width $L_{y}=80a$, $J_{p}$ for the
upper edge is defined as $\sum_{iy=77}^{iy=80}j_{{\bf i}\rightarrow
{\bf i}+\delta x}^{\uparrow}$. With the help of $J_{p}$, one can
quantitatively analyze the local currents. On the whole, the
behavior of the position-related current $J_{p}$ is similar to the
local current configurations, but it is smoother with the position
$x$. More significantly, one can observe from fig. 6(b), the local
currents flow from left to right with rapidly decreasing magnitude
for upper edge, vice versa for the lower edge. This phenomenon can
be attributed to the bulk state assisted backscattering between two
edges. The fact that the position-related current $J_{p}$ in the
bulk is small but nonzero indicates that the bulk states are not
fully localized at the dip. The scattered carriers in the upper edge
can hop through such delocalized bulk states to the lower edge which
leads to the backscattering processes. In other words, disorder not
only destroys the bulk transport but also quickly destroys the edge
transport. Because of this, the conductance $G$ is lower than
plateau value at the dip point. For the region after dip (see fig.
2(a)), increasing of $W$ slowly destroys the bulk transport, but it
also suppresses the bulk-assisted backscattering mechanism at the
same time. It leads to an overall increase of the conductance $G$.
When the bulk states become fully localized, the anomalous plateau
shows up. In fig.6(c), the conductance $G$ versus $W$ with different
sample lengths $L_{x}$ are plotted. The dip feature is clearer for a
large $L_{x}$ because of the increasing probability of the
backscattering between the two edges. However, for all $L_{x}$, the
anomalous plateaus appear with the same disorder strength due to the
fully localized bulk states.

\section{\protect\normalsize CONCLUSIONS}
In summary, the disorder effect in HgTe/CdTe quantum wells is
studied. We confirm the existence of the topological Anderson
insulator(TAI) phase. Conductances calculated for the stripe and
cylinder samples reveal the topological feature of TAI and support
the idea that helical edge states cause the anomalous quantized
plateau. With the help of local-current-vector configurations for
different chemical potentials and disorder strengths, the basic
physical phenomena emerged in the normal QSHE region and in the TAI
region are clearly understood. In particular, the analysis of the
local current configurations provides us with the importance of the
bulk-states-assisted backscattering in TAI that in turn help us to
understand the mechanism of the formation of the disorder induced
edge states.

\section{\protect\normalsize Acknoledgement}
The work is supported by NSFC under Grant Nos. 10525418, 10734110,
and 10821403, and by 973 Program Project No. 2009CB929101. XCX is
supported by US-DOE and Oklahoma C-Spin Center. We thank J. K. Jain,
S.Q. Shen, Xuele Liu and Waigen Zhang for helpful discussions. Part
of the calculations was performed on the HPCC clusters at OSU.

\end{document}